\documentclass[10pt,  twocolumn, twoside, letterpaper, conference, final]{IEEEtran}
\IEEEoverridecommandlockouts
\def\ps@headings{%
\def\@oddhead{\mbox{}\scriptsize\rightmark \hfil \thepage}%
\def\@evenhead{\scriptsize\thepage \hfil \leftmark\mbox{}}%
\def\@oddfoot{}%
\def\@evenfoot{}}
\makeatother
\usepackage{cite}
\usepackage{algorithmicx}
\usepackage[ruled]{algorithm}
\usepackage{algpseudocode}
\usepackage{mathtools}
\DeclarePairedDelimiter\abs{\lvert}{\rvert}

\usepackage{amsmath}
\usepackage{amsfonts}
\usepackage{amssymb}
\DeclareMathOperator*{\argmax}{argmax}
\usepackage{graphicx}
\usepackage{multirow}
\usepackage{textcomp}
\usepackage{diagbox}
\usepackage{colortbl}
\usepackage{xcolor}
\def\BibTeX{{\rm B\kern-.05em{\sc i\kern-.025em b}\kern-.08em
    T\kern-.1667em\lower.7ex\hbox{E}\kern-.125emX}}
\begin{document}

\title{Denial of Service Attacks Detection in Software-Defined Wireless Sensor Networks\\}

\author{\IEEEauthorblockN{Gustavo A. Nunez Segura\IEEEauthorrefmark{1},
Sotiris Skaperas\IEEEauthorrefmark{2}, Arsenia Chorti\IEEEauthorrefmark{3}, 
 Lefteris Mamatas\IEEEauthorrefmark{2} and Cintia Borges Margi\IEEEauthorrefmark{1}}
\IEEEauthorblockA{\IEEEauthorrefmark{1}Escola Polit\'{e}cnica, Universidade de S\~{a}o Paulo, S\~{a}o Paulo, Brazil} 
\IEEEauthorblockA{\IEEEauthorrefmark{2}Department of Applied Informatics, University of Macedonia, Thessaloniki, Greece}
\IEEEauthorblockA{\IEEEauthorrefmark{3}ETIS UMR8051, CY Université, ENSEA, CNRS, F-95000, Cergy, France}
Email: \{gustavoalonso.nunez, cintia\}@usp.br,
        arsenia.chorti@ensea.fr,
        \{sotskap, emamatas\}@uom.edu.gr
}

\maketitle

\begin{abstract}
Software-defined networking (SDN) is a promising technology to overcome many challenges in wireless sensor networks (WSN), particularly  with respect to flexibility and reuse. Conversely, the centralization and the planes' separation turn SDNs vulnerable to new security threats in the general context of distributed denial of service (DDoS) attacks. State-of-the-art approaches to identify DDoS do not always take into consideration restrictions in typical WSNs e.g., computational complexity and power constraints, while further performance improvement is always a target. The objective of this work is to propose a lightweight but very efficient DDoS attack detection approach using change point analysis. Our approach has a high detection rate and linear complexity, so that it is suitable for WSNs. We demonstrate the performance of our detector in software-defined WSNs of 36 and 100 nodes with varying attack intensity (the number of attackers ranges from 5\% to 20\% of nodes). We use change point detectors to monitor anomalies in two metrics: the data packets delivery rate and the control packets overhead.  Our results show that with increasing intensity of attack, our approach can achieve a detection rate close to 100\% and that the type of attack can also be inferred.
\end{abstract}


%%%%%%%%%%%%%%%%%%%%%%%%%%%%%%%%%%%%%%%%%%%%%%%%%%%%%%%%%%%%%%%%%%%%%%%%%%%%%%%%
\section{Introduction} \label{sec:intro}

Software-defined networking (SDN) is a paradigm that was devised to simplify network management, avoid configuration errors and automate infrastructure sharing in wired networks \cite{Ieee2015}. 
The aforementioned benefits motivated the discussion of combining SDN and wireless sensor networks (WSNs) as a solution to many WSN challenges, in particular concerning flexibility and resource reuse \cite{Kobo2017}. This combination is referred to as software-defined wireless sensor networks (SDWSN).
The SDWSN approach decouples the control plane from the data plane and centralizes the control decisions; its main characteristic is the ability to program the network operation dynamically \cite{McKeown2008}. Recent results show that SDWSNs can perform as well as RPL \cite{8805072}. 

On the other hand, the SDN centralization and the planes' separation turn the network vulnerable to new security threats (explained in Section \ref{sec:sdwsnsecurity}), a property that is inadvertently passed on to SDWSNs. Shielding SDNs from these vulnerabilities has already attracted a lot of attention in the literature. 
There are proposals to implement attack detection in Internet of things (IoT) networks using SDN. Sankar and Gurusamy \cite{8215418} proposed \textit{softhings}, an SDN-based IoT framework with security support. The framework was developed for OpenFlow \cite{McKeown2008}, which however, limits its use in networks composed of low-end nodes. The use of support vector machines (SVM) was proposed to detect control plane attacks; it was shown that a detection rate of around 96\% and 98\% could be achieved. The algorithm was tested in Mininet, simulating scenarios with only five 5 nodes and considering one node as attacker.   

Yin \textit{et al.} \cite{8352645} developed the framework SD-IoT, which includes a security system for DDoS attacks detection, based on the difference of packets received by the controller. The difference is calculated using the \textit{cosine similarity} method. This mechanism was devised for networks where all the nodes have periodic communication with the controller, which could be not optimal for very ``restricted'' networks with low-end nodes. Authors tested their proposal through simulations using Mininet. The network size is not explicitly specified, but is inferred to be around 50 to 60 nodes.

Overall, in the case of SDWSNs, due to the resource constraints of the nodes, most of the security mechanisms designed for non-resource constrained SDNs have to be adapted or redesigned. This is one of the major challenges for SDWSN security. Wang \textit{et al.} \cite{WANG2018119} proposed an SDWSN trust management and routing mechanism. They compared their proposal to SDN-WISE when both networks are under attack. The focus of the work is on the selective forwarding attacks and new flow requests. The first attack applies to any type of WSNs, while the second is specific to SDN. The mechanism was tested in simulations with 100 nodes, varying the number of attackers between 5 and 20. Their results show an attack detection rate between 90\% and 96\% when 5 nodes are attackers, and between 60\% and 79\% when 20 nodes are attackers.

Considering the limitations of previous works, our main objective is to propose a mechanism for DDoS detection with, i) a high detection rate, and, ii) low complexity, so that it would be suitable for ``restricted'' networks. To this end, we propose the employment of change of point analysis \cite{skaperas:hal-01997965}\cite{8835019}. We study two DDoS attacks: a false data flow forwarding (FDFF) attack, and a false neighbor information (FNI) attack, chosen to illustrate the proposed algorithm's capabilities in the case of specific SDWSN vulnerabilities that exhibit largely different behavior. Both attacks are explained in Section \ref{sec:dosattacks}. 

% The proposed DDoS attack detector has linear complexity with the network size, is autonomous and non-parametric and exploits state-of-the-art change point analysis, combining an off-line training period with an on-line detector that operates in real-time.

We have tested our approach on the IT-SDN framework\footnote{http://www.larc.usp.br/users/cbmargi/www/it-sdn/} ~\cite{8805072}  and our results show that we can detect these attacks with a detection rate close to 100\%, improving the state of the art; importantly, it is further possible to gain insight regarding the \textit{type of the attack}, based on the metric that provides the quickest detection, a feature, that to the best of our knowledge, breaks new ground in the domain of DDoS analysis for SDWSNs.

% The remaining of this paper is organized as follows. Section \ref{sec:dosattacks} illustrates the FDFF and FNI attacks and their impact on the network performance. In Section \ref{sec:cp}, we present in brief the change point detection algorithm. Experimental methods are presented in Section \ref{sec:methods}. In Section \ref{sec:results} the results are presented and analyzed, while Section \ref{sec:conclusions} concludes the paper.

%%%%%%%%%%%%%%%%%%%%%%%%%%%%%%%%%%%%
\section{Impact of DDoS Attacks in SDWSNs}%on the Network Performance} 

\subsection{SDWSN security analysis} \label{sec:sdwsnsecurity}

The SDN networks security threats are grouped in three sets \cite{7226783}: application plane attacks, control plane attacks, and data plane attacks. Among the three, the control plane attacks are pointed out as the most high impact and attractive \cite{7226783}\cite{Shu2016}, as the control plane is responsible for the overall management of the network \cite{7081073}. This characteristic turns the control plane prone to distributed denial of service (DDoS) attacks. For example, an intruder may flood the network with flow rule requests, which could lead to an exhaustion of the controller's resources. This attack can be intensified using multiple intruders.

The threats and vulnerabilities explained before also apply to SDWSN. Moreover, there are specific attacks that can attain SDWSNs due to resources constraints, for example: in SDWSN the forwarding devices have low storage capacity, which limits the memory assigned for flow tables and buffers. These constraints make the forwarding devices prone to saturation attacks. Also, SDWSN networks are characterized for having a limited bandwidth and low processing power. This means that a saturation attack can also result in a DoS attack.

Another vulnerability concerns the gateway between the SDN controller and the WSN. The gateway has a radio module of limited bandwidth, rendering it a weak link even when the controller has enough resources to overcome an attack. 

For the reasons outlined above, most of the security mechanisms designed for standard SDN networks have to be adapted or redesigned. This is one of the major challenges for SDWSN security.

\label{sec:dosattacks}
%\section{Security analysis} 
\subsection{Impact of DDoS Attacks on Network Performance}
Based on SDWSN specific security vulnerabilities, in a previous work, we studied the impact of three DDoS attacks on SDWSN performance~\cite{OJIOT2019gnunez}. The attacks investigated were: false flow request (FFR), false data flow forwarding (FDFF), and false neighbor information (FNI). 

The FFR attack aimed at increasing the SDWSN controller's processing overhead, as well as the packets' traffic, thus, increasing the number of collisions. Each attacker sent multiple flow rule requests to the controller, while the latter calculated the rule and replied to the request. The impact of the attack was observed to be negligible. The FDFF attack followed the FFR attack main idea of sending false flow rule requests to the controller, however, the execution was based on using each attacker's neighbors (benign nodes). Each attacker sent one data packet to its neighbors tagged with an unknown flow identifier; as the neighbors did not have a rule to apply to the packet, they sent a flow request to the controller asking a rule for the unknown flow identifier. Thus, compared to the FFR, the intensity of the attack was multiplied by the number of neighbors. The FDFF attack tripled the number of control packets in the whole network, but had a minor impact on the delivery rate. For both control and data packets, the delivery rate decreased only between 2\% and 4\%. 

In the FNI attack, each attacker intercepted packets containing neighbor information, modified them with false neighbor information and forwarded them to the controller. The controller updated the network topology graph using the false information, and then reconfigured the network with wrong forwarding rules.
Our main results~\cite{OJIOT2019gnunez} showed that the FNI attack could double the number of control packets in the whole network and had a significant impact on the delivery rate. In the case of the control packets, the delivery rate decreased between 35\% and 50\%. In the case of the data packets, the delivery rate decreased between 20\% and 70\%. 

%Overall, The FFR was considered as an attack with minor impact for the configuration and parameters used. For this reason, in this work we focus on the FDFF and FNI attacks.

%%%%%%%%%%%%%%%%%%%%%%%%%%%%%%%%%%%%
\section{Change Point Detection Algorithm for DDoS}\label{sec:cp}

The study in \cite{OJIOT2019gnunez} provided valuable insight regarding the impact of the FDFF and FNI DDoS attacks on the two metrics under observation, i.e., the mean data packet delivery rate and the mean control packets overhead. Building on this analysis, we formulate the attack detection problem as a hypothesis test, examining whether a change has occurred in the mean value of the time series of the metrics involved. 

In \cite{skaperas:hal-01997965} a change point (CP) detection algorithm was proposed to estimate in real-time the existence, the number, the magnitude and the direction of changes in a time series. To attain these objectives, the algorithm combined (i) off-line and on-line CP schemes; (ii) an improved measurements window segmentation heuristic for the detection of multiple CPs; and (iii) a variation of the moving average convergence divergence (MACD) indicator to detect the direction of changes. The main elements of the algorithm are explained in the following.

\subsection{Basic Off-line Approach} 

The proposed algorithm tests the constancy of the mean values of the time series through a hypothesis test; the null hypothesis is defined as $H_{0}: \mu_{1} = \cdots = \mu_{N}$ against the alternative $H_{1}: \mu_{1} = \cdots = \mu_{k} \ne \mu_{k+1} = \cdots = \mu_{N}$ indicating a change point (CP) at instance $k \in \lbrace 1, N \rbrace$, where $N$ denotes the length the time series and $\mu_i$ the mean value of the time series up to instance $i$. 

Assuming that each sample of the time series $X_1,...,X_N$ can be written as, $X_n=\mu_n+Y_n, \quad1\leqslant{n}\leqslant{N}$, a non-parametric CUSUM test statistic can be developed to identify changes in $\mu$\cite{Aue}; the test statistic can be viewed as a max-type procedure,
\begin{equation}\label{eq:2}
M=\max\limits_{1\leqslant{n}\leqslant{N}}{{C^T_n}\widehat{\Omega}_N^{-1}{C_n}},
\end{equation}
where the parameter $C_n$ is the typical CUSUM,
\begin{equation}\label{eq:3}
C_n=\dfrac{1}{\sqrt{N}}\left(\sum_{i=1}^{n}{X_i}-\dfrac{n}{N}\sum_{i=1}^{N}{X_i}\right),
\end{equation}
and $\widehat{\Omega}_N$ is the estimator of the asymptotic covariance $\Omega$, where
\begin{equation}\label{eq:4}
{\Omega}=\sum_{s=-\infty}^{\infty}{\mathbf{Cov}\left(X_nX_{n-s}\right)}.
\end{equation}

To estimate $\Omega$, %We decided to employ a kernel based estimator instead of a bootstrap based because, considering the low computational time requirement, the first one is more suitable. We employ
the Bartlett estimator was employed \cite{skaperas:hal-01997965}. %, with $k_{BT}$ denoting the Bartlett weight. 
Finally, the critical values for several significance levels $\alpha$ were computed using Monte Carlo simulations. %to approximate the paths of the Brownian bridge on a fine grid. 
The last step is to estimate, if $H_0$ fails, the unknown CP, under $H_1$, given by:
\begin{equation}\label{eq:10}
\widehat{cp}=\dfrac{1}{N}{\argmax_{1\leqslant{n}\leqslant{N}}{M}}.
\end{equation}

\subsection{On-line Phase}
The on-line scheme includes an on-line CUSUM algorithm for the detection of a change in the mean and a MACD indicator to estimate the direction of a change;  $X_{n}$ is expressed as \ref{eq:11}
\begin{equation}\label{eq:11}
X_n= \begin{cases} 
         \mu+Y_n, \hspace{13mm} n=1,\ldots,m+k^*-1 \\
\mu+Y_n+I, \hspace{5mm} n=m+k^*,\ldots 
   \end{cases}
\end{equation}
where $\mu$, $M\in\mathbb{R}$, represent the mean parameters before and after the unknown time of possible change $k^*\in\mathbb{N^*}$ respectively. The term $m$ denotes the length of an initial training period during which there is no change in the mean ($\mu_1=\dots=\mu_m$). In the form of a statistical hypothesis test, the on-line problem is posed as,
\begin{equation}\label{eq:13}
\begin{aligned}
H_0: I=0\\
H_1: I\neq0.
\end{aligned}
\end{equation}
The on-line detection belongs to the category of stopping time procedures, in which for a chosen detector ${TS(m,k)}$ and a given threshold $F(m,k)$ we define the stopping time as: 
\begin{equation}\label{eq:14}
\tau{\left(m\right)}= \begin{cases} 
         \min\lbrace{k\in{\mathbb{N}}: \vert{TS(m,k)}\vert{\geqslant{F(m,k)}}}\rbrace\\
\infty, \text{otherwise} 
   \end{cases} .
\end{equation}
It is necessary to have $\lim_{m\to\infty} P{\left\lbrace\tau_{m}<\infty|H_0\right\rbrace=a}$, 
% \[ \lim_{m\to\infty} P{\left\lbrace\tau_{m}<\infty|H_0\right\rbrace=a},\]
ensuring that the probability of false alarm is asymptotically bounded by $\alpha\in\left(0,1\right)$, and, $\lim_{m\to\infty} P{\left\lbrace{\tau_{m}<\infty|H_1}\right\rbrace=1}$,
% \[ \lim_{m\to\infty} P{\left\lbrace{\tau_{m}<\infty|H_1}\right\rbrace=1},\]
ensuring that under $H_1$ the asymptotic power is unity.
Fulfilling these condition, the threshold $F(m,k)$ was defined as,
\begin{equation}\label{eq:15}
F(m,k)={{c_a}}{g_{\gamma}}{\left({m,k}\right)},
\end{equation}
where the critical value $c_{a}$ is determined from the asymptotic distribution of the detector under $H_0$ and the asymptotic behavior achieved by letting $m\rightarrow\infty$. The weight function,
\begin{equation}\label{eq:16}
g_{\gamma}(m,k)=\sqrt{m}\left(1+\frac{k}{m}\right)\left(\frac{k}{k+m}\right)^\gamma
\end{equation}
depends on the sensitivity parameter $\gamma\in\left[0,1/2\right)$.     
The on-line phase use %two different types of statistics tests: i) 
the standard CUSUM  %and ii) the ratio type CUSUM.
%The standard CUSUM#
detector, given by:
\begin{equation}\label{eq:17}
\Gamma(m,k)=\frac{1}{\widehat{\omega}_{m}}\left(\sum_{i=m+1}^{m+k}{X_i}-\frac{k}{m}{\sum_{i=1}^{m}{X_i}}\right)
\end{equation}
where $\widehat{\omega}_{m}$ denotes the asymptotic variance, that captures the serial dependence between observations.

The corresponding threshold is $F^{\Gamma}(m,k)=c^{\Gamma}_{a}g_{\gamma}(m,k)$ and the critical value is defined as:
\begin{align}
\label{eq:18}
\lim_{m\to\infty} P {\lbrace\tau_{m}<\infty\rbrace} & =\lim_{m\to\infty} P {\left\lbrace\frac{1}{{\widehat{\omega}}_m}{\sup_{1\leqslant{k}\leqslant{\infty}}}{\frac{\abs{\Gamma(m,k)}}{g_{\gamma}(m,k)}}{>c^{\Gamma}_{a}}\right\rbrace}\nonumber \\
&= \left\lbrace\sup_{t\in\left[0,1\right]}{\frac{W(t)}{t^{\gamma}}}>c^{\Gamma}_{a}\right\rbrace=\alpha.
\end{align}

The direction of change is estimated applying the MACD indicator. This indicator is based on an exponential moving average (EMA) filter. More details about this indicator can be found in previous works \cite{skaperas:hal-01997965}\cite{8835019}. 

% \begin{equation}
%     EMA_{p}(n) = \frac{2}{p+1}X_{n} + \frac{p-1}{p+1}EMA_{p}(n-1), 
% \end{equation}
 
% with $p$ denoting the lag parameter. The MACD is derived from the substraction of a short $p_{2}$ lag EMA (sensitive filter) and a longer $p_{3}$ lag EMA (blunt filter), such as shown in Equation \ref{eq:filters1}.

% \begin{equation} \label{eq:filters1}
%     MACD(n) =  EMA_{p_{2}} - EMA_{p_{3}}
% \end{equation}

% Then, the trend indicator (TI) is derived from the substraction of a short $p_{1}$ lag EMA filter of a MACD series and a raw MACD series. The TI indicator expression is shown in Equation \ref{eq:filters2}.

% \begin{equation} \label{eq:filters2}
%     TI(n)  = MACD(n) - EMA_{p_{1}}(MACD(n)),\hspace{0.5cm} p_{1} < p_{2} < p_{3}, 
% \end{equation}

% Finally, we calculate the trend indicator at a time interval ($\hat{cp}_{i}, \hat{cp}_{i}+h$), where $h$ is a threshold parameter. Then, the trend indicator is expressed as shown in Equation \ref{eq:trendmodif}.

% \begin{equation} \label{eq:trendmodif}
%     TI(\hat{cp}_{i}+h) = \sum_{k=\hat{cp}_{i}}^{\hat{cp}_{i}+h}[MACD(X_{k})-EMA_{p1}(MACD(X_{k})]
% \end{equation}

\subsection{Overall algorithm}

Summarizing, the overall algorithm has 5 main steps:
\begin{itemize}
    \item Step 1: define a finite monitoring window $k > 0$ from a starting time instance $m_{s}$,
    \item Step 2: apply the off-line algorithm for the whole historical period $h=\{1, \ldots, m_{s}\}$. If no changes are detected, set $m=h$, conversely, the training sample becomes $m = \{cp_{last},\ldots, m_{s}\}$, where $cp_{last}$ is the last off-line CP detected.
    \item Step 3: apply the on-line procedure $TS(m, k)$ on the interval ${m_{s}, m_{s}+k}$. If an on-line CP ($\hat{cp}^{*}$) is detected, the on-line process stops. Conversely, $\hat{cp}^{*}= 0$, the monitoring ends and proceeds to Step 5.
    \item Step 4: define $k_{cp}=\hat{cp}^{*}$ as a CP and apply the trend indicator. If $TI(k_{cp} > 0)$, announce an upward change. Conversely, announce a downward change.
    \item Step 5: Set a new starting point for the monitoring period. If $k_{cp} > 0$, set $ms = k_{cp} + d$ where $d$ is a constant value defining a period assuming no change, else, set $m_{s} = m_{h}$.
\end{itemize}

\section{Methods} \label{sec:methods}

We employed the CP algorithm in \cite{skaperas:hal-01997965} in SDWSNs under FDFF and FNI attacks. We simulated grid topologies with 36 and 100 nodes, varying the number of attackers in the network (5\% and 20\%).  Each simulation runs during 10 hours and each scenario was replicated 30 times. During the first 8 hours the network operated normally, then the attack is triggered. The choice of 8 hours was made because empirically it was seen that we needed at least 250 samples for the training period and we obtained one sample every 2 minutes.
% The sensor nodes transmitted one data packet every 30 seconds and one management packet every 2 minutes. The management packets contained the information about the network metrics (control packets and data packets delivery rate) \cite{Thamires}.
The simulations were performed using the COOJA simulator \cite{Osterlind2006} and sky motes. The MAC layer was the IEEE 802.15.4, configured to work without radio duty cycle (\verb!nullrdc_driver!). The data sink received the application data, while the management sink received performance metrics information. Notice that the SDN controller is a different node from the sink. Table \ref{tab:parameters} depicts the simulation parameters.
\begin{table}[htb]
\renewcommand{\arraystretch}{1.3}
% if using array.sty, it might be a good idea to tweak the value of
% \extrarowheight as needed to properly center the text within the cells
% \todo{adicionar novos cenários (adicionado tipos de flow setup e P2P-RPL nos MOPs)} 
% \todo{olhar tabela II de \cite{sdnwise-testbed} e ver o que eles colocaram como referência}
\caption{Simulation Parameters}
\label{tab:parameters}
\centering
\resizebox{\columnwidth}{!}{%
\begin{tabular}{ll}
\hline
\multicolumn{2}{|l|}{\textbf{Simulation parameters}}                                                                         \\ \hline
\multicolumn{1}{|l|}{Topology}                          & \multicolumn{1}{l|}{Square grid}                          \\ \hline
\multicolumn{1}{|l|}{Number of nodes}                   & \multicolumn{1}{l|}{36 and 100}     
\\ \hline
\multicolumn{1}{|l|}{Simulation duration}                & \multicolumn{1}{l|}{36000 s}     
\\ \hline
\multicolumn{1}{|l|}{Node boot interval}                & \multicolumn{1}{l|}{$[0, 1]$ s}                           \\ \hline
\multicolumn{1}{|l|}{Number of sinks}                   & \multicolumn{1}{l|}{2}                                 \\ \hline
\multicolumn{1}{|l|}{Sinks position}                    & \multicolumn{1}{l|}{Middle of the grid edge}              \\ \hline
\multicolumn{1}{|l|}{Data traffic rate}                 & \multicolumn{1}{l|}{1 packet every 30 seconds}            \\ \hline
\multicolumn{1}{|l|}{Management traffic rate}           & \multicolumn{1}{l|}{1 packet every two minutes}            \\ \hline
\multicolumn{1}{|l|}{Data payload size}                 & \multicolumn{1}{l|}{10 bytes}                             \\ \hline
\multicolumn{1}{|l|}{Management payload size}           & \multicolumn{1}{l|}{10 bytes}                       \\ \hline
\multicolumn{1}{|l|}{Data traffic start time}           & \multicolumn{1}{l|}{$[2, 3]$ min}                         \\ \hline
\multicolumn{1}{|l|}{Radio module power}                & \multicolumn{1}{l|}{0 dB}                         \\ \hline
\multicolumn{1}{|l|}{Distance between neighbors}        & \multicolumn{1}{l|}{50 m}                         \\ \hline
\multicolumn{1}{|l|}{Attacks begins after}                  & \multicolumn{1}{l|}{28800 s}                         \\ \hline

\\ \hline
\multicolumn{2}{|l|}{\textbf{IT-SDN parameters}}                                                                             \\ \hline
\multicolumn{1}{|l|}{Controller position}               & \multicolumn{1}{l|}{center}                               \\ \hline
% \multicolumn{1}{|l|}{Controller retransmission timeout} & \multicolumn{1}{l|}{60 s}                                 \\ \hline
\multicolumn{1}{|l|}{ND protocol}                       & \multicolumn{1}{l|}{Collect-based}                        \\ \hline
\multicolumn{1}{|l|}{Link metric}                       & \multicolumn{1}{l|}{ETX}                                  \\ \hline
% \multicolumn{1}{|l|}{Neighbor report max frequency}     & \multicolumn{1}{l|}{1 packer per minute}                  \\ \hline
\multicolumn{1}{|l|}{CD protocol}                       & \multicolumn{1}{l|}{none}                                 \\ \hline
\multicolumn{1}{|l|}{Flow setup}                        & \multicolumn{1}{l|}{source routed}                        \\ \hline
\multicolumn{1}{|l|}{Route calculation algorithm}       & \multicolumn{1}{l|}{Dijkstra}                             \\ \hline
\multicolumn{1}{|l|}{Route recalculation threshold}     & \multicolumn{1}{l|}{$10\%$}                               \\ \hline
\multicolumn{1}{|l|}{{Flow setup types}}     & \multicolumn{1}{l|}{{regular or source routed}}                               \\ \hline
\multicolumn{1}{|l|}{{Flow table size}}     & \multicolumn{1}{l|}{{10 entries}}                               \\ \hline
\end{tabular}
}

\end{table}

We analyzed the data packets delivery rate and the control packets overhead. The delivery rate was calculated by dividing the total number of packets successfully received by the total number of packets sent. The control packets overhead was quantified as the total amount of control packets sent. Those metrics were updated every two minutes.

The metrics measuring the performance of the intrusion detection algorithm are: i) detection rate (DR); ii) false positive rate (FPR); iii) false negative rate (FPR); iv) detection time median (DTM), indicating the median of the time instances elapsed from the launch of the attack to the instance it was identified; and v) median absolute deviation (MAD). The detection rate is the ratio between the correctly detected attacks and the total number of attacks. The false positive rate is the ratio between the number of attack events classified as attack and the total number of attack events. The false negative rate is the ratio between attack events classified as non-attack event and the number of attack events. The detection time median is the median of the number of samples required to detect the attack. The median absolute deviation measures the variability of the detection times and is calculated as shown in (\ref{eq:1}), where $X_{i}$ is the detection time for replication $i$, and $\tilde{X}$ is the median of all the detection times, 
\begin{equation} \label{eq:1}
    \mathrm{MAD} = \mathrm{median}(|X_{i} - \tilde{X}|)
\end{equation}

The delivery rate and control overhead time series were analyzed for three monitoring windows and three critical values. We used monitoring periods $K \in \{ 50, 100 , 150\}$ samples. This means that the test statistic is run over $K$ samples to extract changes in the mean value. As critical values we used $\alpha \in \{90\%, 95\%,  99\%\}$. Finally, in this analysis, we discarded the first 15 samples because during this time the network is bootstrapping.

%%%%%%%%%%%%%%%%%%%%%%%%%%%%%%%%%%%%%%%%%%%%%%%%%
\section{Results and Analysis} \label{sec:results}

In this Section we present and analyze the simulation results. In Section \ref{sec:resultsfdff} we compare the FDFF attack detection performance when monitoring the data packets delivery rate and the control overhead. In Section \ref{sec:resultsfni} we repeat this analysis for the FNI attack. 

\subsection{FDFF attack detection} \label{sec:resultsfdff}

Tables \ref{tab:fdff3602} and \ref{tab:fdff10002} summarize the FDFF attack detection results when 5\% of nodes are attackers. The results show that when monitoring the data packets delivery rate, the DR is between 57\% and 73\% for 36 nodes, and between 60\% and 83\% for 100 nodes. The results when monitoring the control packets overhead show two main points: (i) the algorithm has the same detection performance if configured with a monitoring period $K$ of 50 or 150 samples, and (ii) when the monitoring period is configured as $K=100$ samples we obtained a DR between 97\% and 100\%. 

Comparing the FPR and the FNR metrics, we observe that the number of cases classified as false negative is higher than the number of cases classified as false positive. This means, it is more common for the algorithm not to detect a change in the metrics when the network is under attack than to detect a suspicious change in a network without attackers. For example, looking at the results when monitoring the control overhead in Table \ref{tab:fdff3602}, only in one out of nine cases the FPR was different than zero. Conversely, the FNR was different than zero in six of nine cases.

The DTM (detection time median) results show that when monitoring the control packets overhead, the attack detection is faster than when monitoring the delivery rate in all the cases. When monitoring the data packets delivery rate, the DTM is between 31 and 37 samples for 36 nodes, and between 20 and 31 samples for 100 nodes. When monitoring the control packets overhead, the DTM is between 9 and 19 samples for 36 nodes, and between 10 and 19 samples for 100 nodes. The fastest detection is obtained monitoring the control packets overhead using a monitoring period of 100 samples, highlighted in red color.

\begin{table}[t]
\centering
\renewcommand{\arraystretch}{1.3}
\caption{FDFF Attack Detection, 36 Nodes, 5\% Attackers}
\label{tab:fdff3602}
\begin{tabular}{|l|c|c|c|c|c|c|c|c|c|}
%\hline
\multicolumn{10}{c}{\textbf{Data packets delivery rate}}\\ \hline
$K$     & \multicolumn{3}{c|}{\textbf{50}}             & \multicolumn{3}{c|}{\textbf{100}}            & \multicolumn{3}{c|}{\textbf{150}}            \\ \hline
$\alpha$     & \textbf{90} & \textbf{95} & \textbf{99} & \textbf{90} & \textbf{95} & \textbf{99} & \textbf{90} & \textbf{95} & \textbf{99} \\ \hline
\textbf{DTM} & 31           & 33            & 31            & 31           & 37            & 33            & 31           & 31            & 31            \\ \hline
\textbf{MAD}    & 4            & 6             & 4             & 8            & 9             & 10            & 4            & 4             & 4             \\ \hline
\textbf{DR}     & 63         & 67          & 67          & 57         & 70          & 63          & 67         & 73          & 70          \\ \hline
\textbf{FPR}    & 7         & 10          & 7          & 0         & 0          & 0          & 0         & 0          & 0          \\ \hline
\textbf{FNR}     & 30         & 23          & 27          & 43         & 30          & 37          & 33         & 27          &30          \\ \hline
\multicolumn{10}{c}{\textbf{Control overhead}}     \renewcommand{\arraystretch}{1.3}                                                                                      \\ \hline
$K$   & \multicolumn{3}{c|}{\textbf{50}}        & \multicolumn{3}{c|}{\textbf{100}}       & \multicolumn{3}{c|}{\textbf{150}}       \\ \hline
$\alpha$   & \textbf{90} & \textbf{95} & \textbf{99} & \textbf{90} & \textbf{95} & \textbf{99} & \textbf{90} & \textbf{95} & \textbf{99} \\ \hline
\textbf{DTM}   & 19          & 16          & 18          & 12          & \textcolor{red}{9}           &  {11}          & 19          & 16          & 18          \\ \hline
\textbf{MAD} & 3           & 3           & 3           & 3           & 2           & 2           & 3           & 3           & 3           \\ \hline
\textbf{DR}  & 67          & 73          & 67          & 100         & 97          & 100         & 67          & 73          & 67          \\ \hline
\textbf{FPR} & 0           & 0           & 0           & 0           & 3           & 0           & 0           & 0           & 0           \\ \hline
\textbf{FNR}  & 33          & 27          & 33          & 0           & 0           & 0           & 33          & 27          & 33          \\ \hline

\end{tabular}
\end{table}

\begin{table}[t]
\centering
\renewcommand{\arraystretch}{1.3}
\caption{FDFF Attack Detection, 100 nodes, 5\% Attackers}
\label{tab:fdff10002}
\begin{tabular}{|l|c|c|c|c|c|c|c|c|c|}
%\hline
\multicolumn{10}{c}{\textbf{Data packets delivery rate}}\\ \hline 
${K}$      & \multicolumn{3}{c|}{\textbf{50}}             & \multicolumn{3}{c|}{\textbf{100}}            & \multicolumn{3}{c|}{\textbf{150}}            \\ \hline
\textbf{$\alpha$}     & \textbf{90} & \textbf{95} & \textbf{99} & \textbf{90} & \textbf{95} & \textbf{99} & \textbf{90} & \textbf{95} & \textbf{99} \\ \hline
\textbf{DTM} & 24           & 26            & 27          & 22           & 20            & 21          & 29           & 31          & 31            \\ \hline
\textbf{MAD}    & 7          & 6             & 13            & 9            & 10            & 11          & 13           & 9           & 15            \\ \hline
\textbf{DR}     & 60         & 67          &67          & 77         & 83          & 73          & 63         & 67          & 63          \\ \hline
\textbf{FPR}    & 23         & 20          &20          & 10         & 7          & 13          & 0         & 3          & 7          \\ \hline
\textbf{FNR}     &17         & 13          & 13          & 13         & 1           & 13          & 37         & 30          & 30          \\ \hline
\multicolumn{10}{c}{\textbf{Control overhead}}                                                                                           \\ \hline
$K$   & \multicolumn{3}{c|}{\textbf{50}}        & \multicolumn{3}{c|}{\textbf{100}}       & \multicolumn{3}{c|}{\textbf{150}}       \\ \hline
$\alpha$   & \textbf{90} & \textbf{95} & \textbf{99} & \textbf{90} & \textbf{95} & \textbf{99} & \textbf{90} & \textbf{95} & \textbf{99} \\ \hline
\textbf{DTM}   & 19          & 17          & 19          & 13          &  \textcolor{red}{10}          & 12          & 19          & 17          & 19          \\ \hline
\textbf{MAD} & 3           & 3           & 3           & 3           & 2           & 3           & 3           & 3           & 3           \\ \hline
\textbf{DR}  & 60          & 73          & 63          & 100         & 100         & 100         & 60          & 73          & 63          \\ \hline
\textbf{FPR} & 0           & 0           & 0           & 0           & 0           & 0           & 0           & 0           & 0           \\ \hline
\textbf{FNR}  & 40          & 27          & 37          & 0           & 0           & 0           & 40          & 27          & 37          \\ \hline
\end{tabular}
\end{table}

% Tables \ref{} summarize the FDFF attack detection results when 10\% of nodes are attackers. Again, the algorithm obtained the best DR and the fastest detection when monitoring the control packets overhead. In the case of 36 nodes and using a monitoring period of 100 samples, the algorithm obtained a DR of 100\% and DTMs of 6 and 7 samples. The best DR when monitoring the data packets delivery rate was 97\%, obtained for the configuration with 150 samples and all confidence levels. In this case the DTM is between 32 and 35 samples, this means, more than 4 times the best DTM when monitoring the control packets overhead. 

% The scenario with 100 nodes when monitoring the control packets overhead obtained a DR of 100\% when configuring the monitoring period in 50 and 150 samples. In both cases with DTMs of 7 and 8 samples. When configuring the monitoring period in 100 samples, we observed and improvement in the DTM (5 samples), but the DR fell to 90\% and 93\%. 

Tables \ref{tab:fdff3620} and \ref{tab:fdff10020} summarize the FDFF attack detection results when 20\% of nodes are attackers. In the case of 36 nodes, the DR is between 73\% and 83\% when monitoring the data packets delivery rate, and between 87\% and 100\% when monitoring the control packets overhead. In terms of detection time, the best DTM when monitoring the data packets delivery rate was 24 samples and the DTM when monitoring the control packets overhead was 5 samples. %Similar to the observed for the scenarios with 10\% of nodes behaving as attackers, 
Configuring the monitoring period in 100 we obtain the best DTM, but there is a drop in the DR if compared with the cases when using monitoring periods of 50 and 150 samples.

The results for 100 nodes show it is possible to obtain a DR of 100\% monitoring any of the metrics, but there are significant differences in the detection time. The DTM when monitoring the control overhead is between 3 and 4 samples, while when monitoring the data packets delivery rate the DTM is between 7 and 15 samples. Considering the earliest detection with the highest DR for both monitoring metrics, it occurs when using a monitoring period of 100 samples. For both cases the DR obtained was 97\%.
In terms of FPR and FNR, the best performance was obtained when monitoring the control overhead and using a monitoring period of 50 and 150 samples. Monitoring the control overhead using a monitoring window of 100 samples provides a FPR between 3\% and 10\%.

Summarizing, the algorithm is able to detect the FDFF attack using either the data packet packets delivery rate or the control packets overhead as inputs. Notably, the algorithm obtaines a DR of 100\% with both metrics when 20\% of nodes behave as attackers. However, aiming for the quickest detection captured through the detection time median, the algorithm achieved far better results when monitoring the control packets overhead in all scenarios.  %Despite this, the attack detection is faster when monitoring the control packets overhead. 
This is a direct consequence of the type of the attack; the attacker creates multiple flow rule request packets to increase the packet traffic and the controller processing overhead. After some time, the flow table of the nodes around the attacker start to saturate, affecting the data packets delivery rate. This means that the change in the delivery will be detected only after the tables saturation; on the contrary, the number of control packets start to change immediately after the attack is triggered.

%The FDFF attack creates multiple flow rule request packets using unknown flow identifiers to increase the packets traffic and the controller processing overhead. Also, this attack can saturate the 

\begin{table}[t]
\centering
\renewcommand{\arraystretch}{1.3}
\caption{FDFF Attack Detection, 36 nodes, 20\% Attackers}
\label{tab:fdff3620}
\begin{tabular}{|l|c|c|c|c|c|c|c|c|c|}
%\hline
\multicolumn{10}{c}{\textbf{Data packets delivery rate}}\\ \hline 
$K$     & \multicolumn{3}{c|}{\textbf{50}}             & \multicolumn{3}{c|}{\textbf{100}}            & \multicolumn{3}{c|}{\textbf{150}}            \\ \hline
$\alpha$      & \textbf{90} & \textbf{95} & \textbf{99} & \textbf{90} & \textbf{95} & \textbf{99} & \textbf{90} & \textbf{95} & \textbf{99} \\ \hline
\textbf{DTM} & 28           & 28            & 28            & 30         & 24            & 28          & 29           & 28            & 28            \\ \hline
\textbf{MAD}    & 5            & 8             & 6           & 11         & 7             & 8             & 6            & 5             & 8             \\ \hline
\textbf{DR}     & 77         & 80          & 73          & 73         & 83          & 73          & 77         & 80          & 77          \\ \hline
\textbf{FPR}    & 3         & 07          & 7          & 0         & 3          & 0          & 0         & 3          &0          \\ \hline
\textbf{FNR}     & 20         & 13          & 20          & 27         & 13          & 27          & 23         & 17          & 23          \\ \hline
\multicolumn{10}{c}{\textbf{Control overhead}}                                                                                           \\ \hline
$K$    & \multicolumn{3}{c|}{\textbf{50}}        & \multicolumn{3}{c|}{\textbf{100}}       & \multicolumn{3}{c|}{\textbf{150}}       \\ \hline
$\alpha$   & \textbf{90} & \textbf{95} & \textbf{99} & \textbf{90} & \textbf{95} & \textbf{99} & \textbf{90} & \textbf{95} & \textbf{99} \\ \hline
\textbf{M}   & 8           & 7           &  \textcolor{red}{7}            &  \textcolor{red}{5}           & 5           & 5           & 8           &  \textcolor{red}{7}          & 7           \\ \hline
\textbf{MAD} & 2           & 2           & 2           & 1           & 1           & 1           & 2           & 2           & 2           \\ \hline
\textbf{DR}  & 100         & 100         & 100         & 97          & 87          & 97          & 100         & 100         & 100         \\ \hline
\textbf{FPR} & 0           & 0           & 0           & 3           & 13          & 3           & 0           & 0           & 0           \\ \hline
\textbf{FNR}  & 0           & 0           & 0           & 0           & 0           & 0           & 0           & 0           & 0           \\ \hline

\end{tabular}
\end{table}

\begin{table}[t]
\centering
\renewcommand{\arraystretch}{1.3}
\caption{FDFF Attack Detection, 100 nodes, 20\% Attackers}
\label{tab:fdff10020}
\begin{tabular}{|l|c|c|c|c|c|c|c|c|c|}
%\hline
\multicolumn{10}{c}{\textbf{Data packets delivery rate}}\\ \hline 
$K$      & \multicolumn{3}{c|}{\textbf{50}}             & \multicolumn{3}{c|}{\textbf{100}}            & \multicolumn{3}{c|}{\textbf{150}}            \\ \hline
$\alpha$      & \textbf{90} & \textbf{95} & \textbf{99} & \textbf{90} & \textbf{95} & \textbf{99} & \textbf{90} & \textbf{95} & \textbf{99} \\ \hline
\textbf{DTM} & 15         & 13            & 14          & 8          & 7           & 7           & 15           & 14          & 14          \\ \hline
\textbf{MAD}    & 5          & 6           & 5           & 6          & 5           & 5           & 5            & 5           & 5           \\ \hline
\textbf{DR}     & 100         & 93          & 100          & 97         & 93          & 97                      & 100         & 97          & 97          \\ \hline
\textbf{FPR}    & 0         & 7          & 0          & 3         & 7          & 3                      & 0            & 3          & 3          \\ \hline
\textbf{FNR}     & 0         & 0          & 0          & 0         & 0             & 0                          & 0           & 0          & 0          \\ \hline
\multicolumn{10}{c}{\textbf{Control overhead}}                                                                                           \\ \hline
$K$   & \multicolumn{3}{c|}{\textbf{50}}        & \multicolumn{3}{c|}{\textbf{100}}       & \multicolumn{3}{c|}{\textbf{150}}       \\ \hline
$\alpha$    & \textbf{90} & \textbf{95} & \textbf{99} & \textbf{90} & \textbf{95} & \textbf{99} & \textbf{90} & \textbf{95} & \textbf{99} \\ \hline
\textbf{DTM}   & 4           & 4           &  \textcolor{red}{4}           &  \textcolor{red}{3}           & 3           & 3           &  \textcolor{red}{4}           & 4           & 4           \\ \hline
\textbf{MAD} & 0           & 0           & 0           & 0           & 0           & 0           & 0           & 0           & 0           \\ \hline
\textbf{DR}  & 100         & 97          & 100         & 97          & 90          & 97          & 100         & 97          & 100         \\ \hline
\textbf{FPR} & 0           & 3           & 0           & 3           & 10          & 3           & 0           & 3           & 0           \\ \hline
\textbf{FNR}  & 0           & 0           & 0           & 0           & 0           & 0           & 0           & 0           & 0           \\ \hline
\end{tabular}
\end{table}

\subsection{FNI attack detection} \label{sec:resultsfni}

Tables \ref{tab:fni3602} and \ref{tab:fni10002} summarize the FNI attack detection results when 5\% of nodes are attackers. Opposite to the FDFF attack results, the algorithm obtained a better performance detecting the FNI attack when monitoring the data packets delivery rate. In the case of 36 nodes, the DR when monitoring the data packets delivery rate is between 80\% and 93\%, and the DR when monitoring the control packets overhead is between 23\% and 33\%. In the case of 100 nodes, the DR when monitoring the data packets delivery rate is between 83\% and 93\%, and the DR when monitoring the control packets overhead is between 30\% and 70\%. This means, even the best DR when monitoring the control packets overhead is under the worse DR when monitoring the data packets delivery rate. Also, the results show that using a critical value of 90\%, we can obtain a negligible FPR (in our simulation calculated zero).
With respect to the DTM, the best result was obtained by monitoring the data packets delivery rate and the control packets overhead were 6 and 25 samples, respectively. This means the algorithm detects the attack four times faster when monitoring the data packets delivery rate. For 100 nodes, the best DTM when monitoring the data packets delivery rate remains in 6 samples, but when monitoring the control packets overhead it is 29 samples.

\begin{table}[t]
\centering
\renewcommand{\arraystretch}{1.3}
\caption{FNI Attack Detection, 36 nodes, 5\% Attackers}
\label{tab:fni3602}
\begin{tabular}{|l|c|c|c|c|c|c|c|c|c|}
%\hline
\multicolumn{10}{c}{\textbf{Data packets delivery rate}}\\ \hline 
$K$     & \multicolumn{3}{c|}{\textbf{50}}             & \multicolumn{3}{c|}{\textbf{100}}            & \multicolumn{3}{c|}{\textbf{150}}            \\ \hline
$\alpha$      & \textbf{90} & \textbf{95} & \textbf{99} & \textbf{90} & \textbf{95} & \textbf{99} & \textbf{90} & \textbf{95} & \textbf{99} \\ \hline
\textbf{DTM} & 7            & 6           & 7           & 8          & 7           &  \textcolor{red}{6}            & 7            & 6             & 6             \\ \hline
\textbf{MAD}    & 3            & 4           & 3           & 4          & 3           & 3             & 2            & 4           & 4           \\ \hline
\textbf{DR}     & 93         & 83          & 93          & 93         & 80          & 93          & 93         & 83          & 87          \\ \hline
\textbf{FPR}    & 0         & 10          & 0          & 0         & 13          & 0          & 0         & 10          & 7          \\ \hline
\textbf{FNR}     & 7         & 7          & 7          & 7         & 7          & 7          & 7         & 7          & 6          \\ \hline
\multicolumn{10}{c}{\textbf{Control overhead}}                                                                                           \\ \hline
$K$   & \multicolumn{3}{c|}{\textbf{50}}        & \multicolumn{3}{c|}{\textbf{100}}       & \multicolumn{3}{c|}{\textbf{150}}       \\ \hline
$\alpha$    & \textbf{90} & \textbf{95} & \textbf{99} & \textbf{90} & \textbf{95} & \textbf{99} & \textbf{90} & \textbf{95} & \textbf{99} \\ \hline
\textbf{DTM}   & 28          & 25          & 27          & 35          & 26          & 33          & 28          & 25          & 27          \\ \hline
\textbf{MAD} & 6           & 7           & 9           & 4           & 3           & 5           & 6           & 7           & 9           \\ \hline
\textbf{DR}  & 27          & 33          & 27          & 20          & 27          & 23          & 27          & 33          & 27          \\ \hline
\textbf{FPR} & 3           & 3           & 3           & 0           & 0           & 0           & 0           & 0           & 0           \\ \hline
\textbf{FNR}  & 70          & 63          & 70          & 80          & 73          & 77          & 73          & 67          & 73          \\ \hline

\end{tabular}
\end{table}

\begin{table}[t]
\centering
\renewcommand{\arraystretch}{1.3}
\caption{FNI Attack Detection, 100 nodes, 5\% Attackers}
\label{tab:fni10002}
\begin{tabular}{|l|c|c|c|c|c|c|c|c|c|}
%\hline
\multicolumn{10}{c}{\textbf{Data packets delivery rate}}\\ \hline 
$K$      & \multicolumn{3}{c|}{\textbf{50}}             & \multicolumn{3}{c|}{\textbf{100}}            & \multicolumn{3}{c|}{\textbf{150}}            \\ \hline
$\alpha$     & \textbf{90} & \textbf{95} & \textbf{99} & \textbf{90} & \textbf{95} & \textbf{99} & \textbf{90} & \textbf{95} & \textbf{99} \\ \hline
\textbf{DTM} & 6            &  \textcolor{red}{6}           & 6             & 6            & 6             & 6             & 6            & 6             & 6             \\ \hline
\textbf{MAD}    & 4          & 4           & 3             & 3          & 3             & 2             & 4            & 4             & 4             \\ \hline
\textbf{DR}     & 87         & 93          & 83          & 83         & 83          & 83          & 83         & 90          & 87          \\ \hline
\textbf{FPR}    & 13         & 7          & 17          & 17         & 17          & 17          & 13         & 10          & 13          \\ \hline
\textbf{FNR}     & 0         & 0          & 0          & 0         & 0          & 0          & 3         & 0          & 0          \\ \hline
\multicolumn{10}{c}{\textbf{Control overhead}}                                                                                           \\ \hline
$K$  & \multicolumn{3}{c|}{\textbf{50}}        & \multicolumn{3}{c|}{\textbf{100}}       & \multicolumn{3}{c|}{\textbf{150}}       \\ \hline
$\alpha$    & \textbf{90} & \textbf{95} & \textbf{99} & \textbf{90} & \textbf{95} & \textbf{99} & \textbf{90} & \textbf{95} & \textbf{99} \\ \hline
\textbf{DTM}   & 34          & 29          & 33          & 35          & 37          & 37          & 34          & 29          & 33          \\ \hline
\textbf{MAD} & 7           & 7           & 7           & 10          & 7           & 8           & 7           & 8           & 8           \\ \hline
\textbf{DR}  & 63          & 70          & 67          & 30          & 47          & 37          & 63          & 70          & 67          \\ \hline
\textbf{FPR} & 0           & 0           & 0           & 0           & 0           & 0           & 0           & 0           & 0           \\ \hline
\textbf{FNR}  & 37          & 30          & 33          & 70          & 53          & 63          & 37          & 30          & 33          \\ \hline
\end{tabular}
\end{table}

% Tables \ref{} summarize the FNI attack detection results when 10\% of nodes are attackers. For 36 nodes the algorithm obtained a DR of 100\% in all the configurations when monitoring the data packets delivery rate. Conversely, the best DR obtained when monitoring the control packets overhead was 70\%. About the DTM, the number of samples required to detect the attack when monitoring control packets overhead remains in 25, the same value obtained when 5\% of nodes are attacker. In the case of monitoring the data packets delivery rate, the best DTM obtained was 4 samples. This means, 2 samples less than the best DTM when 5\% of nodes are attackers.

% Similar results are present in the topology of 100 nodes. The algorithm obtained a DR between 97\% and 100\%, and a DTM between 6 and 9 samples when monitoring the data packets delivery rate. In the cases monitoring the control packets overhead, the DR is between 60\% and 87\% and the DTM is between 34 and 40 samples. Comparing the results for both cases, the algorithm obtained has a better detection accuracy when monitoring the data packets delivery rate and is, at least, 3.77 times faster than the cases monitoring the control packets overhead. Also, we observed the worst result for both cases, considering DR and DTM, was obtained when the monitoring period was set in 100 samples. This applies for 36 and 100 nodes

Lastly, Tables \ref{tab:fni3620} and \ref{tab:fni10020} summarize the FNI attack detection results when 20\% of nodes are attackers. For 36 nodes, the results remain similar to the case of 5\% of nodes are attackers. In the case of 100 nodes, the DR when monitoring the data packets delivery rate is between 97\% and 100\%, and the DR when monitoring the control packets delivery rate is between 93\% and 97\%. 
About the DTM, the results for the scenarios when monitoring the data packets delivery rate are between 4 and 9 samples. The results for this same metric when monitoring the control packets overhead are between 24 and 26 samples. This means, for grid topologies with 100 nodes where 20\% of nodes are attackers, we obtain similar DRs regardless of the monitoring metric, but when monitoring the delivery rate the detection is at least 3 times faster.

Summarizing our findings, the algorithm is able to detect the FNI attack monitoring either the data packet packets delivery rate or the control packets overhead. Then, comparing the detection performance based on the detection rate and the detection time median, the algorithm obtained a far better performance when monitoring the data packets delivery rate in all scenarios. %The only case where the algorithm did not obtain a DR of 100\%, considering all the possible configurations, was when 5\% nodes were attackers. In this case the highest DR obtained was 93\% and the DTM was 7 samples. 
This effect is directly related to the type of the attack; in the FNI attack, the attackers intercept the control packets that contain neighbor information, modify them, and then forward them to the controller. This means this attack can lead to a network misconfiguration using few control packets. %This behavior coincides with the DR and DTM results.  
% The results obtained when 10\% of nodes behave as attackers follow similar trends. Tables \ref{tab:fni3610} and \ref{tab:fni10010} in Appendix \ref{app:1} summarize the main results.

\begin{table}[t]
\centering
\renewcommand{\arraystretch}{1.3}
\caption{FNI Attack Detection, 36 nodes, 20\% Attackers}
\label{tab:fni3620}
\begin{tabular}{|l|c|c|c|c|c|c|c|c|c|}
%\hline
\multicolumn{10}{c}{\textbf{Data packets delivery rate}}\\ \hline 
$K$      & \multicolumn{3}{c|}{\textbf{50}}             & \multicolumn{3}{c|}{\textbf{100}}            & \multicolumn{3}{c|}{\textbf{150}}            \\ \hline
$\alpha$      & \textbf{90} & \textbf{95} & \textbf{99} & \textbf{90} & \textbf{95} & \textbf{99} & \textbf{90} & \textbf{95} & \textbf{99} \\ \hline
\textbf{DTM} &  \textcolor{red}{7}             &  \textcolor{red}{7}             &  \textcolor{red}{7}             & 7            & 7             & 7             & 8          &  \textcolor{red}{7}             &  \textcolor{red}{7}            \\ \hline
\textbf{MAD}    & 2          & 2             & 2             & 3            & 4           & 3             & 2            & 2             & 2             \\ \hline
%\textbf{DR}     & \multicolumn{9}{c|}{100} \\ \hline
%\textbf{FPR}     & \multicolumn{9}{c|}{0} \\ \hline
%\textbf{FNR}     & \multicolumn{9}{c|}{0} \\ \hline
\textbf{DR}     & 100         & 100          & 100          & 100         & 100          & 100          & 100         & 100          & 100          \\ \hline
\textbf{FPR}    & 0         & 0          & 0          & 0         & 0          & 0          & 0         & 0          & 0          \\ \hline
\textbf{FNR}      & 0         & 0          & 0          & 0         & 0          & 0          & 0         & 0          & 0           \\ \hline
\multicolumn{10}{c}{\textbf{Control overhead}}                                                                                           \\ \hline
$K$   & \multicolumn{3}{c|}{\textbf{50}}        & \multicolumn{3}{c|}{\textbf{100}}       & \multicolumn{3}{c|}{\textbf{150}}       \\ \hline
$\alpha$    & \textbf{90} & \textbf{95} & \textbf{99} & \textbf{90} & \textbf{95} & \textbf{99} & \textbf{90} & \textbf{95} & \textbf{99} \\ \hline
\textbf{DTM}   & 26          & 24          & 26       & 26          & 24          & 27          & 26          & 24          & 26        \\ \hline
\textbf{MAD} & 8           & 7           & 7           & 17          & 11          & 13          & 8           & 7           & 7           \\ \hline
\textbf{DR}  & 57          & 70          & 60          & 43          & 63          & 57          & 57          & 70          & 60          \\ \hline
\textbf{FPR} & 0           & 0           & 0           & 0           & 0           & 0           & 0           & 0           & 0           \\ \hline
\textbf{FNR}  & 43          & 30          & 40          & 57          & 37          & 43          & 43          & 30          & 40          \\ \hline

\end{tabular}
\end{table}

\begin{table}[t]
\centering
\renewcommand{\arraystretch}{1.2}
\caption{FNI Attack Detection, 100 nodes, 20\% Attackers}
\label{tab:fni10020}
\begin{tabular}{|l|c|c|c|c|c|c|c|c|c|}
%\hline
\multicolumn{10}{c}{\textbf{Data packets delivery rate}}\\ \hline
$K$       & \multicolumn{3}{c|}{\textbf{50}}             & \multicolumn{3}{c|}{\textbf{100}}            & \multicolumn{3}{c|}{\textbf{150}}            \\ \hline
$\alpha$      & \textbf{90} & \textbf{95} & \textbf{99} & \textbf{90} & \textbf{95} & \textbf{99} & \textbf{90} & \textbf{95} & \textbf{99} \\ \hline
\textbf{DTM} & 9          & 10           & 10           &  \textcolor{red}{8}           & 9           &  \textcolor{red}{8}             & 10          & 12            & 11            \\ \hline
\textbf{MAD}    & 5            & 8           & 7           & 4            & 6             & 4             & 5          & 9             & 8             \\ \hline
\textbf{DR}     & 100         & 100          & 100          & 100         & 100          & 100          & 100         & 100          & 97          \\ \hline
\textbf{FPR}    & 0         & 0          & 0          & 0         & 0          & 0          & 0         & 0          & 3          \\ \hline
\textbf{FNR}     & 0         & 0          & 0          & 0         & 0          & 0          & 0         & 0          & 0          \\ \hline
\multicolumn{10}{c}{\textbf{Control overhead}}                                                                                           \\ \hline
$K$  & \multicolumn{3}{c|}{\textbf{50}}        & \multicolumn{3}{c|}{\textbf{100}}       & \multicolumn{3}{c|}{\textbf{150}}       \\ \hline
$\alpha$   & \textbf{90} & \textbf{95} & \textbf{99} & \textbf{90} & \textbf{95} & \textbf{99} & \textbf{90} & \textbf{95} & \textbf{99} \\ \hline
\textbf{DTM}   & 27          & 24          & 26          & 26          & 25          & 25          & 27          & 24          & 26          \\ \hline
\textbf{MAD} & 6           & 3           & 6           & 6           & 6           & 6           & 6           & 3           & 6           \\ \hline
\textbf{DR}  & 93          & 97          & 97          & 93          & 97          & 93          & 93          & 97          & 97          \\ \hline
\textbf{FPR} & 0           & 0           & 0           & 0           & 0           & 0           & 0           & 0           & 0           \\ \hline
\textbf{FNR}  & 7           & 3           & 3           & 7           & 3           & 7           & 7           & 3           & 3           \\ \hline
\end{tabular}
\end{table}

%%%%%%%%%%%%%%%%%%%%%%%%%%%%%%%%%%%%%%%%%%%%%%%%%%%
 \section{Conclusions and Future Work} \label{sec:conclusions}
SDWSNs are exposed to new security threats that may not affect traditional WSNs. Recent proposals in the literature for the identification of DDoS attacks in SDWSN do not always consider ``restricted'' networks and there is also the need to improve the solutions' performance.
In this work we provide a solution for DDoS attack detection for SDWSN. We identify an attack by monitoring changes in the mean values of two metrics, the network data packets delivery rate and the control packets overhead. To detect a change in either metric due to a DDoS attack, we use state-of-the-art non-parametric and on-line change point detection algorithms \cite{skaperas:hal-01997965}. We performed experiments for two SDWSN DDoS attacks, in topologies of 36 and 100 nodes, and with varying number of attackers. The attacks implemented were the FDFF and the FNI. 

Our results showed that it is feasible to detect those attacks by monitoring either the data packets delivery rate or control packets metrics. However, targeting the quickest detection possible, far superior detection performance was achieved for the FDFF when monitoring the control packets overhead. Conversely, results showed a far better performance in detecting the FNI attack when monitoring the data packets delivery rate.  In either cases, the detection rate increased to even 100\% with increasing attack intensity, while the agility of the detection is noteworthy, with either attack identified within $3-10$ samples from its launch. Notably, different metrics have been shown to be better indicators for different types of attack, \textit{allowing to detect  not only the existence,  but, potentially the type of the attack}. 

As the detector's algorithmic complexity is linear to the size of the network and the number of metrics monitored, the proposed approach could scale to include other metrics. 
In future work, we will test the algorithm monitoring the change of other network metrics to see if we can improve the detection performance. We would like to test the algorithm analyzing the metrics by regions or clusters to obtain more information about the attacker location. Also, we will repeat the experiments reducing the simulator Tx/Rx success ratio. 
\vspace{-0.1 cm}
\section*{Acknowledgments}\vspace{-0.1 cm}
This study was financed in part by the Coordena\c{c}\~{a}o de Aperfei\c{c}oamento de Pessoal de N\'{i}vel Superior - Brasil (CAPES) - Finance Code 001 and by the ELIOT project (ANR-18-CE40-0030 / FAPESP 2018/12579-7).
Gustavo A. Nunez Segura is supported by Universidad de Costa Rica. 
\vspace{-0.1 cm}
\bibliographystyle{IEEEtran}
\bibliography{bibliography}

\begin{thebibliography}{10}
\providecommand{\url}[1]{#1}
\csname url@samestyle\endcsname
\providecommand{\newblock}{\relax}
\providecommand{\bibinfo}[2]{#2}
\providecommand{\BIBentrySTDinterwordspacing}{\spaceskip=0pt\relax}
\providecommand{\BIBentryALTinterwordstretchfactor}{4}
\providecommand{\BIBentryALTinterwordspacing}{\spaceskip=\fontdimen2\font plus
\BIBentryALTinterwordstretchfactor\fontdimen3\font minus
  \fontdimen4\font\relax}
\providecommand{\BIBforeignlanguage}[2]{{%
\expandafter\ifx\csname l@#1\endcsname\relax
\typeout{** WARNING: IEEEtran.bst: No hyphenation pattern has been}%
\typeout{** loaded for the language `#1'. Using the pattern for}%
\typeout{** the default language instead.}%
\else
\language=\csname l@#1\endcsname
\fi
#2}}
\providecommand{\BIBdecl}{\relax}
\BIBdecl

\bibitem{Ieee2015}
D.~{Kreutz}, F.~M.~V. {Ramos}, P.~E. {Veríssimo}, C.~E. {Rothenberg},
  S.~{Azodolmolky}, and S.~{Uhlig}, ``{Software-Defined Networking: A
  Comprehensive Survey},'' \emph{Proc. IEEE Proc.}, vol. 103, no.~1, pp.
  14--76, Jan 2015.

\bibitem{Kobo2017}
H.~I. {Kobo}, A.~M. {Abu-Mahfouz}, and G.~P. {Hancke}, ``{A Survey on
  Software-Defined Wireless Sensor Networks: Challenges and Design
  Requirements},'' \emph{IEEE Access}, vol.~5, pp. 1872--1899, 2017.

\bibitem{McKeown2008}
N.~McKeown, T.~Anderson, H.~Balakrishnan, G.~Parulkar, L.~Peterson, J.~Rexford,
  S.~Shenker, and J.~Turner, ``{OpenFlow: Enabling Innovation in Campus
  Networks},'' \emph{SIGCOMM Comput. Commun. Rev.}, vol.~38, no.~2, pp. 69--74,
  Mar. 2008.

\bibitem{8805072}
R.~C.~A. {Alves}, D.~A.~G. {Oliveira}, G.~A. {Nunez Segura}, and C.~B. {Margi},
  ``{The Cost of Software-Defining Things: A Scalability Study of
  Software-Defined Sensor Networks},'' \emph{IEEE Access}, vol.~7, pp.
  115\,093--115\,108, Aug 2019.

\bibitem{8215418}
S.~S. {Bhunia} and M.~{Gurusamy}, ``{Dynamic attack detection and mitigation in
  IoT using SDN},'' in \emph{27th Int. Telecommun. Netw. and Appl. Conf.
  (ITNAC)}, Nov 2017, pp. 1--6.

\bibitem{8352645}
D.~{Yin}, L.~{Zhang}, and K.~{Yang}, ``{A DDoS Attack Detection and Mitigation
  With Software-Defined Internet of Things Framework},'' \emph{IEEE Access},
  vol.~6, pp. 24\,694--24\,705, 2018.

\bibitem{WANG2018119}
R.~Wang, Z.~Zhang, Z.~Zhang, and Z.~Jia, ``{ETMRM: An Energy-efficient Trust
  Management and Routing Mechanism for SDWSNs},'' \emph{Computer Networks},
  vol. 139, pp. 119 -- 135, 2018.

\bibitem{skaperas:hal-01997965}
S.~Skaperas, L.~Mamatas, and A.~Chorti, ``{Early Video Content Popularity
  Detection with Change Point Analysis},'' in \emph{{IEEE Global Commun. Conf.
  (GLOBECOM)}}, Abhu-Dhabi, United Arab Emirates, Dec. 2018.

\bibitem{8835019}
S.~{Skaperas}, L.~{Mamatas}, and A.~{Chorti}, ``{Real-Time Video Content
  Popularity Detection Based on Mean Change Point Analysis},'' \emph{IEEE
  Access}, vol.~7, pp. 142\,246--142\,260, 2019.

\bibitem{7226783}
I.~Ahmad, S.~Namal, M.~Ylianttila, and A.~Gurtov, ``{Security in Software
  Defined Networks: A Survey},'' \emph{IEEE Commun. Surveys Tuts.}, vol.~17,
  no.~4, pp. 2317--2346, Fourthquarter 2015.

\bibitem{Shu2016}
Z.~Shu, J.~Wan, D.~Li, J.~Lin, A.~V. Vasilakos, and M.~Imran, ``{Security in
  Software-Defined Networking: Threats and Countermeasures},'' \emph{Mobile
  Netw. and Appl.}, vol.~21, no.~5, pp. 764--776, Oct 2016.

\bibitem{7081073}
A.~Akhunzada, E.~Ahmed, A.~Gani, M.~K. Khan, M.~Imran, and S.~Guizani,
  ``{Securing software defined networks: taxonomy, requirements, and open
  issues},'' \emph{IEEE Commun. Mag.}, vol.~53, no.~4, pp. 36--44, April 2015.

\bibitem{OJIOT2019gnunez}
G.~A.~N. Segura, C.~B. Margi, and A.~Chorti, ``{Understanding the Performance
  of Software Defined Wireless Sensor Networks Under Denial of Service
  Attack},'' \emph{Open Journal of Internet Of Things (OJIOT)}, 2019, special
  Issue: Proc. Int. Workshop Very Large Internet of Things (VLIoT 2019) in
  conjunction with the VLDB 2019 Conf. Los Angeles, United States.

\bibitem{Aue}
A.~Aue and L.~Horváth, ``Structural breaks in time series,'' \emph{Journal of
  Time Series Analysis}, vol.~34, no.~1, pp. 1--16, 2013.

\bibitem{Osterlind2006}
F.~{Osterlind}, A.~{Dunkels}, J.~{Eriksson}, N.~{Finne}, and T.~{Voigt},
  ``{Cross-Level Sensor Network Simulation with COOJA},'' in \emph{Proc. IEEE
  Conf. Local Comput. Netw. (LCN)}, Nov 2006, pp. 641--648.

\end{thebibliography}

\end{document}